\documentclass[twocolumn,amsmath,amssymb,10pt,superscriptaddress,a4paper,letterpaper,fleqn]{revtex4-1}
\usepackage{amssymb}
\usepackage{epsfig}
\usepackage{graphicx}
\usepackage{dcolumn}
\usepackage{array}
\usepackage{bm}
\usepackage{fancyheadings}
\usepackage{longtable}
\usepackage{multirow}
\usepackage{float}
\usepackage{afterpage}  
\usepackage{color}

\setlongtables
\usepackage[breaklinks=true,linkbordercolor={1 1 1}]{hyperref}

\begin{document}
\setcounter{page}{1}

\title{
%% Please do not remove the line below
\qquad \\ \qquad \\ \qquad \\  \qquad \\  \qquad \\ \qquad \\ 
%% Change title if necessary
Light-ion production from O, Si, Fe and Bi induced by 175~MeV quasi-monoenergetic neutrons}

\author{R.~Bevilacqua}
\thanks{Present address: European Commission -- Joint Research Centre -- Institute for Reference Materials and Measurements, B-2440 Geel, Belgium}
\email[]{Riccardo.BEVILACQUA@ec.europa.eu}
\author{S.~Pomp}
\email[]{Stephan.Pomp@physics.uu.se}
\author{K.~Jansson}
\author{C.~Gustavsson}
\author{M.~\"Osterlund}
\author{V.~Simutkin}
\affiliation{Department of Physics and Astronomy, Uppsala University, 751 20 Uppsala, Sweden}

\author{M.~Hayashi}
\author{S.~Hirayama}
\author{Y.~Naitou}
\author{Y.~Watanabe}
\affiliation{Department od advanced Energy Engineering Science, Kyushu University, 816-8580 Fukuoka, Japan}

\author{A.~Hjalmarsson}
\author{A.~Prokofiev}
\affiliation{The Svedberg Laboratory, Uppsala University, 751 21 Uppsala, Sweden}

\author{U.~Tippawan}
\affiliation{Chiang Mai University, 50200 Chiang Mai, Thailand}

\author{F.-R.~Lecolley}
\author{N.~Marie}
\affiliation{LPC, Universit\'e de Caen, 14050 Caen, France}

\author{S.~Leray}
\author{J.-C.~David}
\affiliation{CEA, Centre de Saclay, Irfu/SPhN, F-91191 Gif-sur-Yvette, France}

\author{S.~Mashnik}
\affiliation{Los Alamos National Laboratory, Los Alamos, NM 87545, USA}

\date{\today} 
%\received{8 March 2013; revised received XX June 2013; accepted XX September 2013}

\begin{abstract}
{
We have measured double-differential cross sections in the interaction of 175~MeV quasi-monoenergetic neutrons with O, Si, Fe and Bi. We have compared these results with model calculations with INCL4.5-Abla07, MCNP6 and TALYS-1.2. We have also compared our data with PHITS calculations, where the pre-equilibrium stage of the reaction was accounted respectively using the JENDL/HE-2007 evaluated data library, the quantum molecular dynamics model (QMD) and a modified version of QMD (MQMD) to include a surface coalescence model. The most crucial aspect is the formation and emission of composite particles in the pre-equilibrium stage.
}
\end{abstract}
\maketitle

\lhead{Light-ion production $\dots$}
%\chead{NUCLEAR DATA SHEETS}
\rhead{R.~Bevilacqua \textit{et al.}}
\lfoot{}
\rfoot{}
\renewcommand{\footrulewidth}{0.4pt}

\section{INTRODUCTION}

Accelerator-driven incineration of nuclear waste may play an important role in future scenarios both in countries investing in the nuclear renaissance and in countries deciding to phase out nuclear energy. However, development of accelerator-driven systems requires new and more reliable neutron-induced cross-section data in the intermediate region from 20 to 200~MeV. In this energy region the entire structural properties of the target nucleus may participate in the nuclear reaction, and the nuclear data community expressed the need to obtain benchmark experimental data to improve present models and evaluated nuclear data libraries~\cite{HPRL}.

A series of neutron induced cross-section measurements at 96~MeV have been conducted at the The Svedberg Laboratory (TSL), Uppsala~\cite{Pomp}. However, experimental data are still scarce above 100~MeV. In this perspective, we have measured cross-sections for light-ion production in the interaction of 175 MeV quasi-monoenergetic neutrons (QMN) with C, O, Si, Fe, Bi and U. 

In this work, we compare O, Si, Fe and Bi experimental data with INCL4.5-Abla07~\cite{INCL,ABLA}, MCNP6~\cite{MCNP6} using CEM03.03~\cite{Mashnik}, TALYS-1.2~\cite{TALYS} and PHITS~\cite{PHITS} model calculations. While most models are able to reproduce the production of protons, the most crucial aspect is to describe the formation and emission of composite particles in the pre-equilibrium stage. 

\section{MATERIALS AND METHODS}

The experimental data presented in this work were measured with the Medley setup~\cite{NIMA} at TSL. Here a quasi-monoenergetic pulsed neutron beam is available, with maximum peak energy of 175~MeV. Time-of-flight information was used to reduce the contribution of the low-energy component of the neutron spectrum, as described in Ref.~\cite{NIMA}. Model calculations presented in this paper were folded with the experimental neutron spectrum, for comparison with measured data.

The Medley spectrometer is an array of eight three-elements $\Delta$E-$\Delta$E-E telescope detectors designed to measure protons, deuterons, tritons, $^3$He and $\alpha$-particles at eight angles in the laboratory system from 20$^{\circ}$ to 160$^{\circ}$. The present configuration of Medley allows good particle separation and at the same time a low identification threshold (2~MeV for protons) and a wide dynamic range (up to 170~MeV). Experimental details and data analysis procedures are described in~Ref.~\cite{NIMA}.

Determination of the experimental cross sections was achieved measuring elastic (n,p) scattering from a CH$_2$ target at 20$^{\circ}$; counts from other reaction targets where then normalized to the same neutron flux and the same number of scattering centers.

\section{RESULTS and DISCUSSION}

\begin{figure}[!t]
\includegraphics[width=0.90\columnwidth]{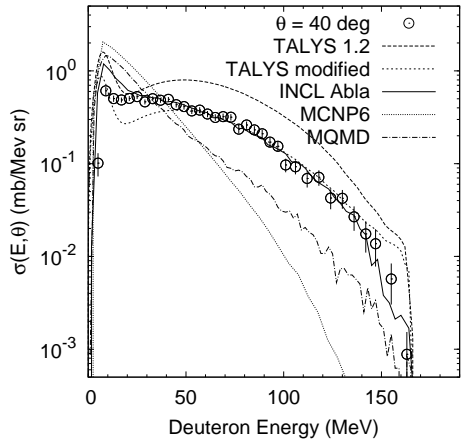}
\caption{Experimental deuteron production from Fe at 40$^{\circ}$ in comparison with model calculations.}
\label{fig1}
\end{figure}

\begin{figure}[!t]
\includegraphics[width=0.90\columnwidth]{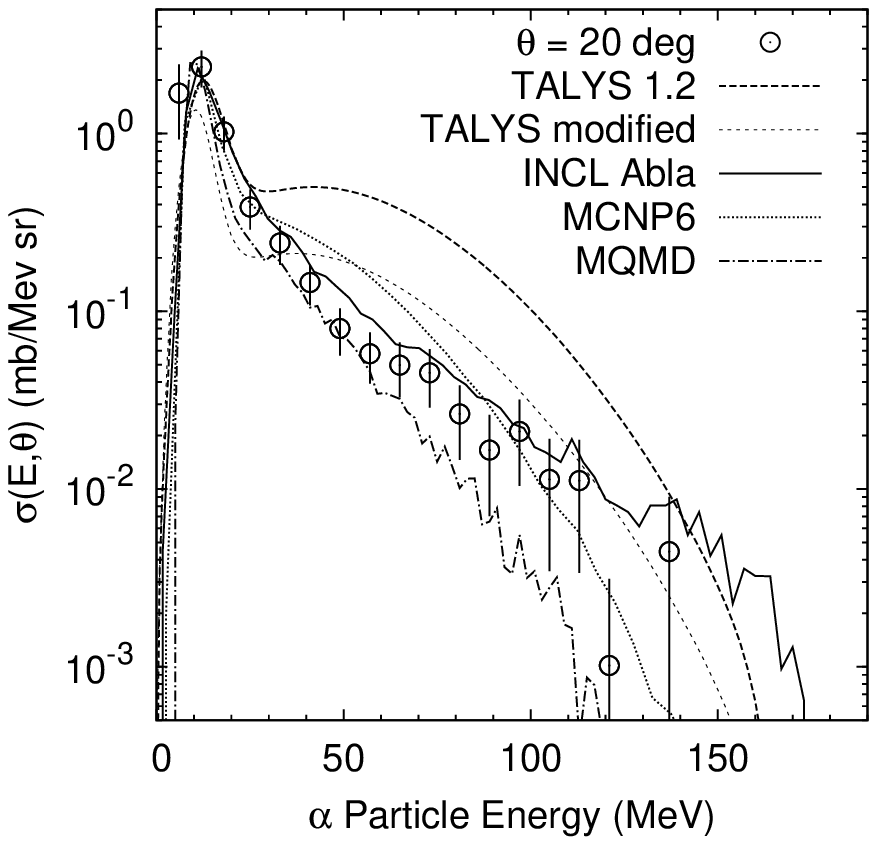}
\caption{Experimental $\alpha$~particle production from Fe at 20$^{\circ}$ in comparison with model calculations}
\label{fig2}
\end{figure}

\begin{figure}[!t]
\includegraphics[width=0.90\columnwidth]{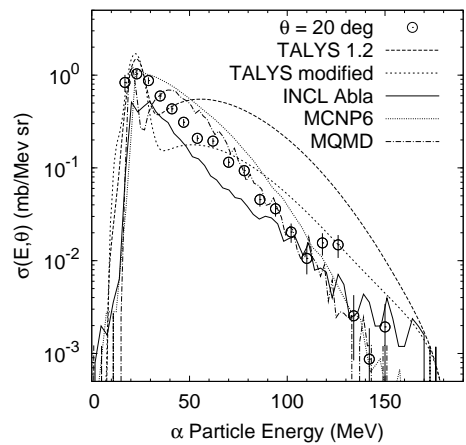}
\caption{Experimental $\alpha$~particle production from Bi at 20$^{\circ}$ in comparison with model calculations}
\label{fig3}
\end{figure}

To describe the pre-equilibrium emission of composite light-ions, TALYS complements the two-component exciton model with direct-like mechanisms, according to the Kalbach systematics~\cite{Kalbach}. We observe how TALYS overestimates the production of deuterons (Fig.~\ref{fig1}) and $\alpha$~particles (Fig.~\ref{fig2},~\ref{fig3}). We observe the same trend also in the pre-equilibrium production of other composite light-ions, at all emission angles. We may improve the predictive capabilities of TALYS by reducing the contribution of the nucleon transfer mechanism in the pre-equilibrium process (TALYS-modified)~\cite{modTALYS}. This result was achieved introducing an energy dependent scaling on the {\it cstrip} parameter in TALYS. 

However, since the TALYS code does not include multiple pre-equilibrium emission of composite particles, the TALYS-modified model calculations underestimate the cross section at low energies in the pre-equilibrium stage of the reaction (Fig.~\ref{fig1},~\ref{fig2},~\ref{fig3}).

The INCL4.5-Abla07 calculations integrates the intra-nuclear cascade (INC) model~\cite{INCL} with the ABLA de-excitation model~\cite{ABLA}. For Fe and Bi targets, INCL-Abla07 well describes the pre-equilibrium deuteron production (e.g. Fig.~\ref{fig1}) and it is in good agreement with production of other composite light-ions (e.g. Fig.~\ref{fig2},~\ref{fig3}).  The INC model is generally used to describe reactions involving neutron energies above 200~MeV. However, our results show how the INC model may be applied to describes reactions with neutrons of lower energies.

MCNP6~\cite{MCNP6} uses the Cascade-Exciton Model (CEM)~\cite{Mashnik} as implemented in the code CEM03.03 as its default event-generator. CEM assumes that nuclear reactions occur generally in three stages. The first stage is the INC. When the cascade stage of a reaction is complete, CEM uses the coalescence model to ÒcreateÓ  high-energy d, t, $^3$He, and $\alpha$~particles via final-state interactions among emitted cascade nucleons, already outside of the target. The second stage is an improved version of the modified exciton model of pre-equilibrium decay followed by the third equilibrium evaporation/fission stage of the reaction. But if the residual nuclei after the INC have atomic numbers with A$<$13, CEM uses the Fermi breakup model to calculate their further disintegration instead of using  the pre-equilibrium and evaporation models. We see that MCNP6 calculations reproduce the  emission of $\alpha$~particles (Fig.~\ref{fig2},~\ref{fig3}), but underestimates the forward emission of other light ions (e.g.~Fig.~\ref{fig1}).

\begin{figure}[!h]
\includegraphics[width=0.90\columnwidth]{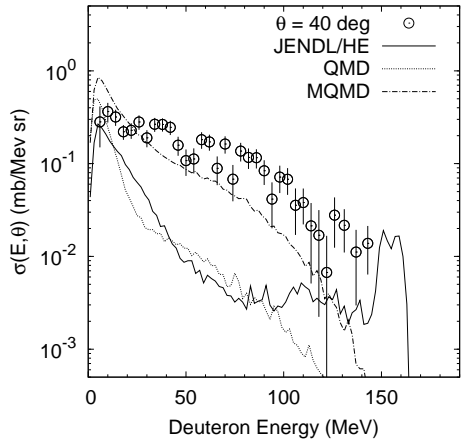}
\caption{Experimental deuteron production from O at 40$^{\circ}$ in comparison with model calculations}
\label{fig4}
\end{figure}

\begin{figure}[!h]
\includegraphics[width=0.90\columnwidth]{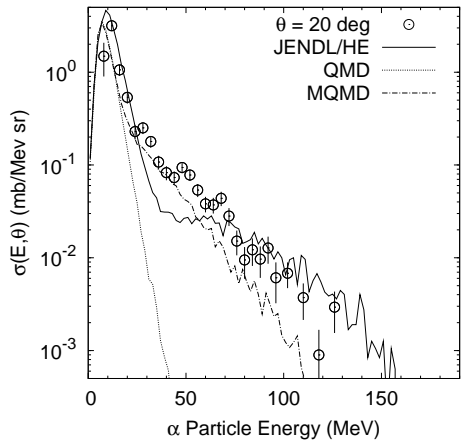}
\caption{Experimental $\alpha$~particle production from Si at 20$^{\circ}$ in comparison with model calculations}
\label{fig5}
\end{figure}

Finally, we calculated double differential cross sections with the PHITS code~\cite{PHITS}, applying three different methods to account for the pre-equilibrium stage of the reactions: using evaluated data from the JENDL/HE-2007 library~\cite{JENDL}, the quantum molecular dynamics (QMD) model and a modified QMD (MQMD)~\cite{JQMD} to include a surface coalescence model. In all three cases, the statistical decay was calculated with the generalized evaporation model.  Whereas production of protons is well described by all calculations, the QMD model cannot account for the pre-equilibrium production of composite ions (Fig.~\ref{fig4},~\ref{fig5}). The inclusion of a coalescence model in the QMD model improves the results for all composite ions (e.g.~Fig.~\ref{fig4},~\ref{fig5}), however the underestimation of pre-equilibrium deuterons at forward angles (Fig.~\ref{fig1},~\ref{fig4}) seems to indicate the need to introduce direct-like reaction mechanisms in the pre-equilibrium stage.

The discrepancies between PHITS calculations with data from the JENDL/HE-2007 library and experimental production of deuterons (Fig.~\ref{fig4}), tritons and $^3$He, indicate the need to improve the evaluation of these reaction data.\\

\section{CONCLUSIONS}

We have measured double-differential cross sections for several elements in the interaction with 175 MeV quasi monoenergetic neutrons, and we have compared our results with model calculations.  

TALYS consistently overestimates the pre-equilibrium emission of all composite light-ions from Fe and Bi. We showed that reducing the contribution of direct-like production mechanisms, according to the Kalbach systematics, improves TALYS prediction capabilities. MCNP6 describes better the emission of $\alpha$~particles from Fe and Bi than of other composite ions, which are consistently underestimated in particular at forward emission angles. We have also seen that introducing a coalescence mechanism in the QMD model improves the predictive power of the PHITS code, and gives good agreement with $\alpha$~particles data. Both MQMD and MCNP6 results seem to suggest the need to introduce direct-like mechanisms to account for the forward emission of deuterons, tritons and $^3$He. Finally, INCL4.5-Abla07 describes fairly well the production of deuterons from Fe and Bi, but discrepancies emerge for other composite light-ions.

\section*{ACKNOWLEDGEMENTS}

R.B. is grateful to the ENEN Association and to Uppsala University for the financial support to participate in the ND2013 Conference. U.T. expresses his gratitude to the Thailand Research Fund (TRF) for financial support under Project No. MGR5280165.

\end{document}